\documentclass[10pt]{article}
\usepackage{a4,epsfig,amsmath,amssymb,latexsym,euscript,oldgerm,rotating,
            picinpar}
\begin{document}
\begin{center}
\centering{\large\bf Thermodynamical implications of a protein model with water interactions}\\

  \vspace*{1cm}
  \centering{Audun Bakk\footnote{Corresponding author.} and Johan S.\ H{\o}ye\\
 {\it Department of Physics, Norwegian University of Science and 
     Technology, NTNU, N-7491 Trondheim, Norway}\\
  \vspace*{0.5cm}
  Alex Hansen\footnote{Permanent Address: Department of Physics, Norwegian University of Science and Technology, NTNU, N-7491 Trondheim, Norway.} and Kim Sneppen\\
  {\it NORDITA and Niels Bohr Institute, Blegdamsvej 17, DK-2100 Copenhagen, Denmark}}\\
  \vspace*{0.5cm}
  \centering{(July 5, 2000)}
\end{center}
\vspace*{0.5cm}

\begin{abstract}
We refine a protein model that reproduces fundamental aspects of protein thermodynamics. The model exhibits two transitions, hot and cold unfolding. The number of relevant parameters is reduced to three: 1) binding energy of folding relative to the orientational energy of bound water, 2) ratio of degrees of freedom between the folded and unfolded protein chain and 3) the number of water molecules that can access the hydrophobic parts of the protein interior. By increasing the number of water molecules in the model, the separation between the two peaks in the heat capacity curve comes closer, which is more consistent with experimental data. In the end we show that if we, as a speculative assumption, assign only two distinct energy levels for the bound water molecules, we obtain better correspondence with experiments.  
\end{abstract}

PACS: 05.70.Ce, 87.14.Ee, 87.15.Cc 

\section{Introduction}
\label{sec:1}
Proteins are crucial components in all living organisms. In order to have biological functionality at physiological temperatures it is important that they have an exclusively ordered state, termed the {\it native} state. Anfinsen showed that the native state is {\it genetically} determined\,\cite{Anfinsen73}, which means that each protein, with its specific amino acid sequence, folds into an unique conformation. The experiment by Anfinsen also proved that the native state is thermodynamically determined, {\it i.e.}\hspace{0.13cm}the state in which Gibbs free energy of the whole system is lowest. It is now commonly accepted that folding of the polypeptide chain is {\it thermodynamically} driven\,\cite{Makhatadze95}.

A peculiar feature of proteins is that they fold on time scales from  
$10^{-3}$ s to\\
 $1$ s. If one calculates the folding time of this process simply by taking the folding as stochastic, one finds astronomical time scales\,\cite{Levinthal68}. This is called the ``Levinthal paradox''. A resolution of this apparent paradox is outlined in a recent review by Shakhnovich\,\cite{Shakhnovich99}, where he discribes how the protein forms at first a ``nucleation-condensate''\,\cite{Fersht95,Itzhaki95} via thermal fluctuations of the polypeptide chain, whereupon a transition state (TS) occurs, carrying common features to the native state, in which the protein descends downhill in the Gibbs free energy landscape to the native state. The recent point of view is that the ``TS-pathway'' is not a concrete mechanistical pathway, on which every position corresponds to a unique conformation. Instead  a ``statistical pathway'' is introduced, where a new step forward on the pathway means reaching a more favorable statistical ensemble of conformations with regard to Gibbs free energy. However, every step along the path, each describing an ensemble of conformations, should have some common structural features which acts like checkpoints for the folding. Further these checkpoints of increasing order is likely to follow a {\it folding pathway}\,\cite{Baldvin99,Hansen98a,Hansen99,Hansen98b,Bakk00}, where one particular point on the pathway depends on the assumption that the main structures of the earlier steps are conserved. 

Unfolding of the polypeptide chain by increasing the temperature is somewhat intuitive, but what is rather surprising is that proteins unfolds at low temperatures, {\it i.e.}\hspace{0.13cm}they become denaturated and not biological functional. {\it Cold} denaturation seems to be a general property of small globular proteins\,\cite{Privalov90,Chen89}. 

The paper is organized as follows. In Sec.\,\ref{sec:2} we present the model and calculate the partition function. In Sec.\,\ref{sec:3} we discuss the thermodynamics of the model, and present results for folding and unfolding transitions.

\section{The physical model}
\label{sec:2}

\subsection{Polypeptide chain}
\label{sub:21}
\begin{figure}
   \caption{}
   \vspace{2cm}
   \centering{\epsfig{figure=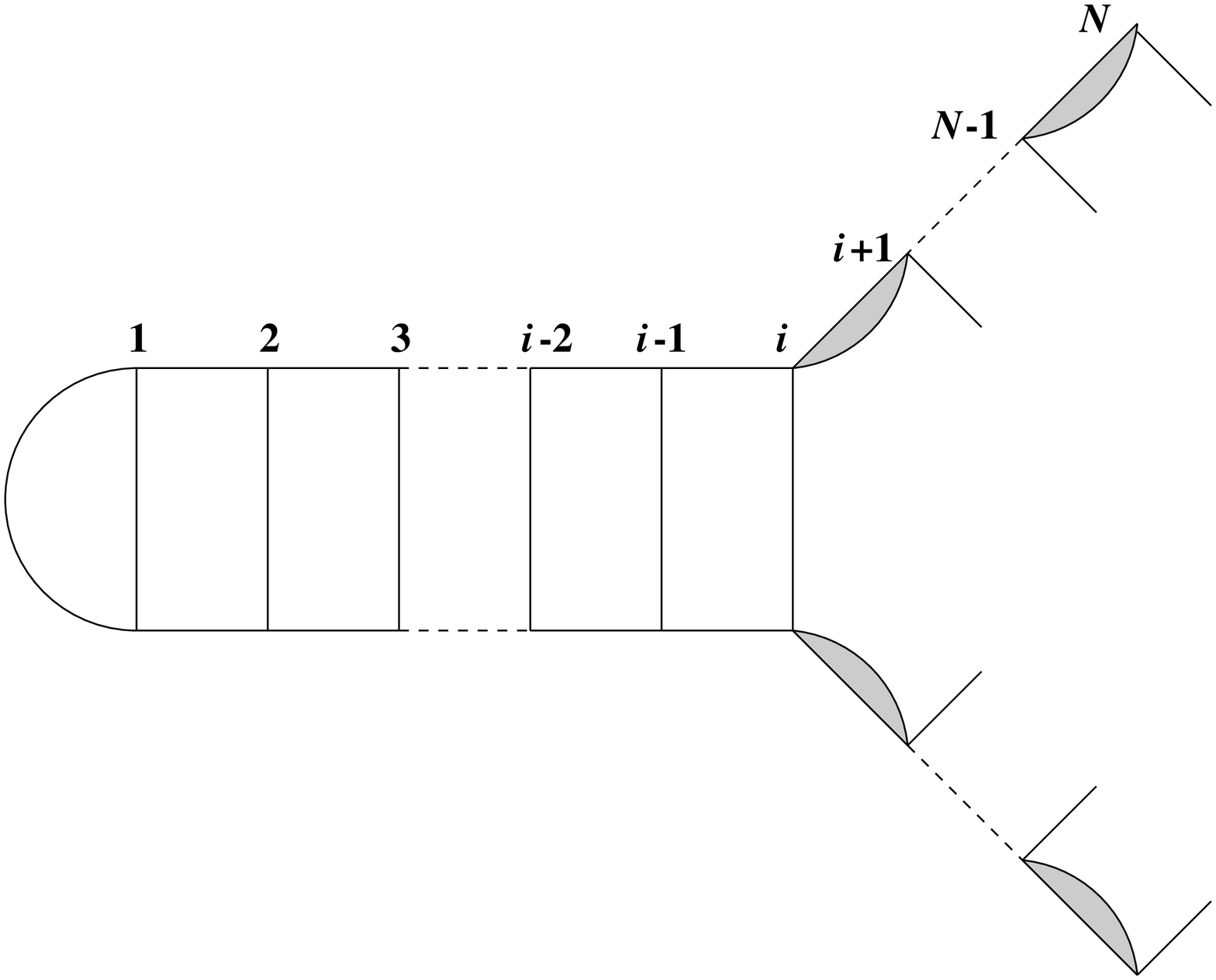,width=\linewidth}}
   \label{fig:1}
\end{figure}

We refine a physical model for a small globular protein, which builds on earlier models by Hansen {\it et al.}\,\cite{Hansen98a,Hansen99} and Bakk {\it et al.}\,\cite{Bakk00}. The protein is viewed as a zipper (Fig.\,\ref{fig:1}), in analogy to the model of Dill {\it et al.}\,\cite{Dill93}, which is a 1-dimensional model of a folding pathway. The complex 3-dimensional protein is equipped with $N$ contact points which we here call {\it nodes}. Each individual node is assigned an energy of $-\epsilon_0 <0$ if it is folded (native), zero otherwise\,\cite{Bryngelson87,Bryngelson90}. This means that a folded node is energetically favorable. Requiring that if node {\it i} is folded, all nodes $j<i$ are also folded, is an implication of the pathway. The point of view that the nodes are distinct contact points in space is a simplification. Folding node 1 means finding the ``nucleation-condensate'', which is reached through a condensation down to a structure which marks the beginning of the folding pathway, and guides the protein into the native state. Each individual node is regarded as statistical ensembles due to the previous discussion in Section\,\ref{sec:1}, and they are likely to form non-local contacts which may be important for the cooperativity\,\cite{Shakhnovich99,Abkevich95,Privalov96}. However, the specific nodes do have some common structural motifs. What the specific mechanism forming this ``nucleation-condensate'' is not considered in this paper, but we {\it assume} that the condensate exists and restrict the study to the TS-pathway, that eventually folds the protein into its native conformation. 

We introduce binary contact variables $\phi_i\in\{ 0,1\}$. $\phi_i=0$ means that node $i$ is open (unfolded), and $\phi_i=1$ means that node $i$ is folded. Assuming $N$ nodes, a Hamiltonian ($H_1$) for the energies associated to the polypeptide chain is in a compact way written\,\cite{Hansen98a,Hansen99,Hansen98b,Bakk00}       

\begin{equation}
    H_1=-\epsilon_0(\phi_1 + \phi_1\phi_2  
           + \cdots+\phi_1\phi_2\cdots\phi_N)\quad .
    \label{H1}
  \end{equation}

\noindent
Product terms ($\phi_1\cdots\phi_i$) meet the assumption about a folding pathway, because if $\phi_i=0$, all terms containing $j\geq i$ vanish.

The unfolded protein will access some more degrees of freedom relative to the native protein, because an unfolded polypeptide backbone will have rotational freedom represented by the {\it dihedral} angles\,\cite{Creighton93}. This can be further simplified to one ``pseudodihedral'' angle\,\cite{Peticolas80}, and is incorporated  by assigning each single unfolded node $f$ degrees of freedom. The parameter $f$ is interpreted as the relative increase in the degrees of freedom for an unfolded node compared to a folded node.   

\subsection{Water interactions}
Introduction of water is important for several reasons. First, proteins is {\it in vivo} exposed to water and second, water has several peculiar properties due to the polarity of, and the hydrogen bonds between water molecules. Makhatadze and Privalov\,\cite{Makhatadze95} states that in sum hydration effects destabilize the native state, and decreasing temperature implies increasing destabilizing action. This is termed as the ``hydrophobic force'', and the water-protein interaction is incorporated by an energy {\it ladder} representing each individual water molecule associated to the unfolded parts of the protein ({\it i.e.} all nodes where $\phi_i =0$)\,\cite{Hansen98a,Hansen99,Bakk00} 

\begin{equation}
  \omega_{ij}=
    \begin{cases}
      -\varepsilon_w +(g-1)\delta\\
      \hspace{0.828cm}\vdots\\
      -\varepsilon_w +2\delta\\
      -\varepsilon_w +\delta\\
      -\varepsilon_w\quad .
    \end{cases}
  \label{ladder}
\end{equation}

\noindent
$\omega_{ij}$ is the energy for water molecule $j$ at node $i$. $\varepsilon_w>0$. Interactions between the water molecules are not considered in this paper. Eq.\,\ref{ladder} is interpreted as all available energies for water molecule associated to the unfolded node $i$. Here we will let $M$ water molecules be associated to each unfolded node, whereas Hansen {\it et al.}\,\cite{Hansen98a,Hansen99} and Bakk {\it et al.}\,\cite{Bakk00} restricted this number to one. No water is supposed to access a folded node, {\it i.e.} the protein interior. 

The ladder contains $g$ equidistant energies which give an entropy contribution while node $i$ is folded, because then the water is unbounded. Hence a folded node implies an entropy contribution from $g^M$ degrees of freedom. The ladder is of course a simplification, and is connected to the need of some sort of energy levels to make it energerically favorable to unfold at low temperatures. Thus the energy ladder in Eq.\,\ref{ladder} is introduced for computational convenience. We note that the proposed energy ladder in fact is nothing but the quantized energy levels of a magnetic dipole in an external magnetic field. However, in the limit $g\rightarrow\infty$ (with $g\,\delta$ finite), the classical limit for a magnetic moment of a fixed length is obtained. This in turn is equivalent to an electric dipole in an electric field. The latter can be interpreted as a direct physical model of dipolar water molecules that feel an effective electric field from the protein. In a protein, an electrical field arises from the permanent and induced charges on the protein surface that becomes exposed after unfolding of a node. This field will interact with the nearest water molecules (dipoles) and structure them. The quantitative aspects of the folding problem will probably need a discussion of additional interactions, but this will not be considered here. Fig.\,\ref{fig:1} is a schematic illustration of a partly folded protein containing some water associated to the hydrophobic parts that uncovers upon unfolding of the nodes.  

By using the same notation as in Eq.\,\ref{H1}, the energy associated to water-protein interactions $H_2$, becomes

\begin{equation}
  \begin{split}
    H_2=&(1-\phi_1)(\omega_{11}+\omega_{12}+\cdots +\omega_{1M})
        +(1-\phi_1\phi_2)(\omega_{21}+\omega_{22}+\cdots +\omega_{2M})\\
        &+\cdots +(1-\phi_1\phi_2\cdots\phi_N)(\omega_{N1}
                     +\omega_{N2}+\cdots+\omega_{NM})\quad .
    \label{H2}
  \end{split}
\end{equation}

\subsection{The partition function}
The Hamiltonian $H=H_1+H_2$ describing the entire system is then
\begin{equation}
  \begin{split}
    H=&-\epsilon_0(\phi_1 + \phi_1\phi_2 
           + \cdots+\phi_1\phi_2\cdots\phi_N)\\
          &+(1-\phi_1)(\omega_{11}+\omega_{12}+\cdots +\omega_{1M})
           +(1-\phi_1\phi_2)(\omega_{21}+\omega_{22}+\cdots +\omega_{2M})\\
          &+\cdots +(1-\phi_1\phi_2\cdots\phi_N)(\omega_{N1}
                     +\omega_{N2}+\cdots+\omega_{NM})\quad .
\end{split}
\end{equation}
The partition function $Z=\sum_{i=0}^{N}Z_i$\,, where the term $Z_i$ corresponds to folding of all nodes $\leq i$ (pathway assumption), becomes
\begin{equation}
  \label{Z_i}
  Z_i=f^{N-i}\, g^{iM}\, e^{i\epsilon_0\beta}
       \left( e^{\varepsilon_w\beta}
       \frac{1-e^{-g\delta\beta}}{1-e^{-\delta\beta}}\right)^{(N-i)M}\quad .
\end{equation}   
$\beta\equiv 1/T$ is a rescaled inverse absolute temperature where the Boltzmann constant is absorbed in $T$. $Z_0$ means that all nodes are open, {\it i.e.} a complete unfolded protein. 
The factor $f^{N-i}$ in Eq.\,\ref{Z_i} arises from the degrees of freedom in the polypeptide chain that are available in the $N-i$ unfolded nodes. Further the product term $g^{iM}$ is the entropy deliberated from $M$ free non-interacting water molecules associated to $i$ folded nodes. $e^{i\epsilon_0\beta}$ is the Boltzmann factor from $i$ contact energies $-\epsilon_0$ in the polypeptide chain. The last term in brackets is simply the sum over all distinct levels in one water-ladder raised to the power of the number of water molecules $(N-i)M$,  bounded to the unfolded hydrophobic parts of the protein.    
A rearrangement of Eq.\,\ref{Z_i} gives
\begin{equation}
  \label{Z_i2}
  Z_i=\left ( g^M\, e^{\epsilon_0\beta}\right )^N\, r^{i-N}\quad ,
\end{equation}
where we have defined 
\begin{equation}
  \label{r1}
  r\equiv \left[ \frac{g}{f^{1/M}}\, e^{(\epsilon_0/M-\varepsilon_w)\beta}\,
            \frac{1-e^{-\delta\beta}}{1-e^{-g\delta\beta}}
           \right]^M \quad .
\end{equation}
We put or assume that $\delta\beta\ll 1$ ({\it i.e.} $g\rightarrow\infty$), which means an infinite small level spacing in the water ladder. Hence a Taylor expansion yields $1-e^{-\delta\beta}\,\approx\,\delta\beta$ and Eq.\,\ref{r1} can be rewritten into
\begin{equation}
  \label{r2}
  r\, =\,  \left[ a\, e^{\mu\beta}\frac{\beta}{\sinh{\beta}}\right]^M \quad .
\end{equation}
$a\equiv 1/f^{1/M}$ and the inverse temperature is rescaled by $g\delta\beta/2\, \rightarrow\, \beta$. The parameter $a$ reflects the ratio of the degrees of freedom between the folded and the unfolded units of protein chain. The new energy parameter 
$\mu\equiv (\epsilon_0/M -\varepsilon_w+g\delta/2)\, /\, (g\delta/2)$ is proportional to the binding energy of each node, and may be interpreted as an {\it effective chemical potential} for each single protein. 
Changing the environments of the protein, {\it i.e.}\hspace{0.13cm}adding denaturants or changing pH, changes this chemical potential. 

We calculate the partition function by simply summing up the $Z_i$ terms in Eq.\,\ref{Z_i2}
\begin{equation}
  \label{Z}
  Z\, =\,\sum_{i=0}^N Z_i\, =\, g^{NM}\, e^{c\beta}\, \frac{1}{r^N}\sum_{i=0}^N r^i\, =\, g^{NM}\, e^{c\beta}
\, \frac{1-r^{-(N+1)}}{1-r^{-1}}\quad ,
\end{equation}  
where $c\equiv 2N\epsilon_0\,/\,g\delta$.

The {\it order parameter} (``reaction coordinate'') in this system is $n$, which is the degree of folding, {\it i.e.} the mean of the number of folded nodes divided by $N$. 
\begin{equation}
  \label{n}
  n=\frac{\sum_{i=0}^{N}i\, Z_i}{Z}
   =\frac{1}{M}\,\frac{\partial}{\partial(\mu\beta)}\,(\ln{Z})
   =\frac{r}{N}\, \frac{N\,r^{N+1}-(N+1)\,r+1}{(1-r^{N+1})\,(1-r)}\quad .
\end{equation}

\section{Thermodynamical calculations and discussion}
\label{sec:3}

\subsection{Continuum limit of the water energy levels}
\label{sub:31}
The heat capacity is 
$C=\beta^2\cdot \partial^2 (\ln{Z})/\partial\beta^2$. This function is independent of the prefactor $g^{MN}\,e^{c\beta}$ in $Z$. Furthermore, $Z$ contains the function $r$, which has only three parameters; the amplitude factor $a$, the effective chemical potential $\mu$ and the number of water molecules per unfolded node $M$. We assume that the number of nodes is a constant, let us say $N=100$, reflecting a typical number of residues in a small protein. The number of {\it relevant} parameters in our physical model is now reduced from the initial six: $f, g, \epsilon_0, \varepsilon_w, \delta$ and $M$, to only three parameters: $a, \mu$ and $M$. 

\begin{figure}
   \caption{}
   \vspace{2cm}
   \centering{\epsfig{figure=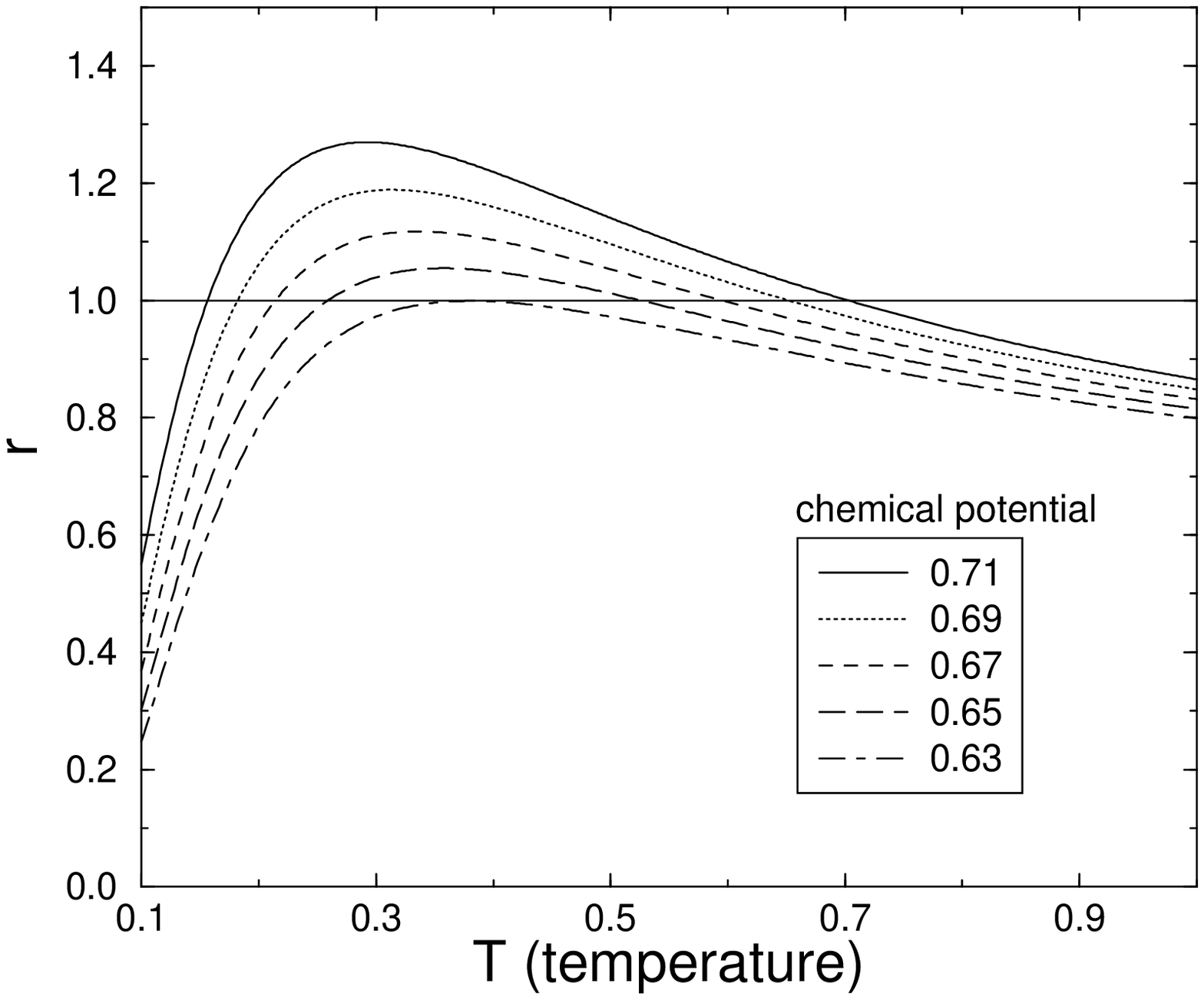,width=\linewidth}}
   \label{fig:2}
\end{figure}

The partition function in Eq.\,\ref{Z}, and thus the heat capacity $C$, is apparently most sensitive to changes in $r$ for values $r\approx 1$. The function $r$ is plotted in Fig.\,\ref{fig:2} for $a=0.5$ and $M=1$. We see the effect of an decreasing effective chemical potential, by the decreasing separation of the two intersections for $r=1$. Larger $M$ implies only a smaller and higher function $r$, while the intersections for $r=1$ is independent of the specific value of $M$. $\mu_c \approx 0.63$ is a critical effective chemical potential, and $\mu <\mu_c$ makes the protein denaturated at all temperatures. This critical point was studied for $M=1$ in Ref\,\cite{Hansen99}.        

\begin{figure}
   \caption{}
   \vspace{2cm}
   \centering{\epsfig{figure=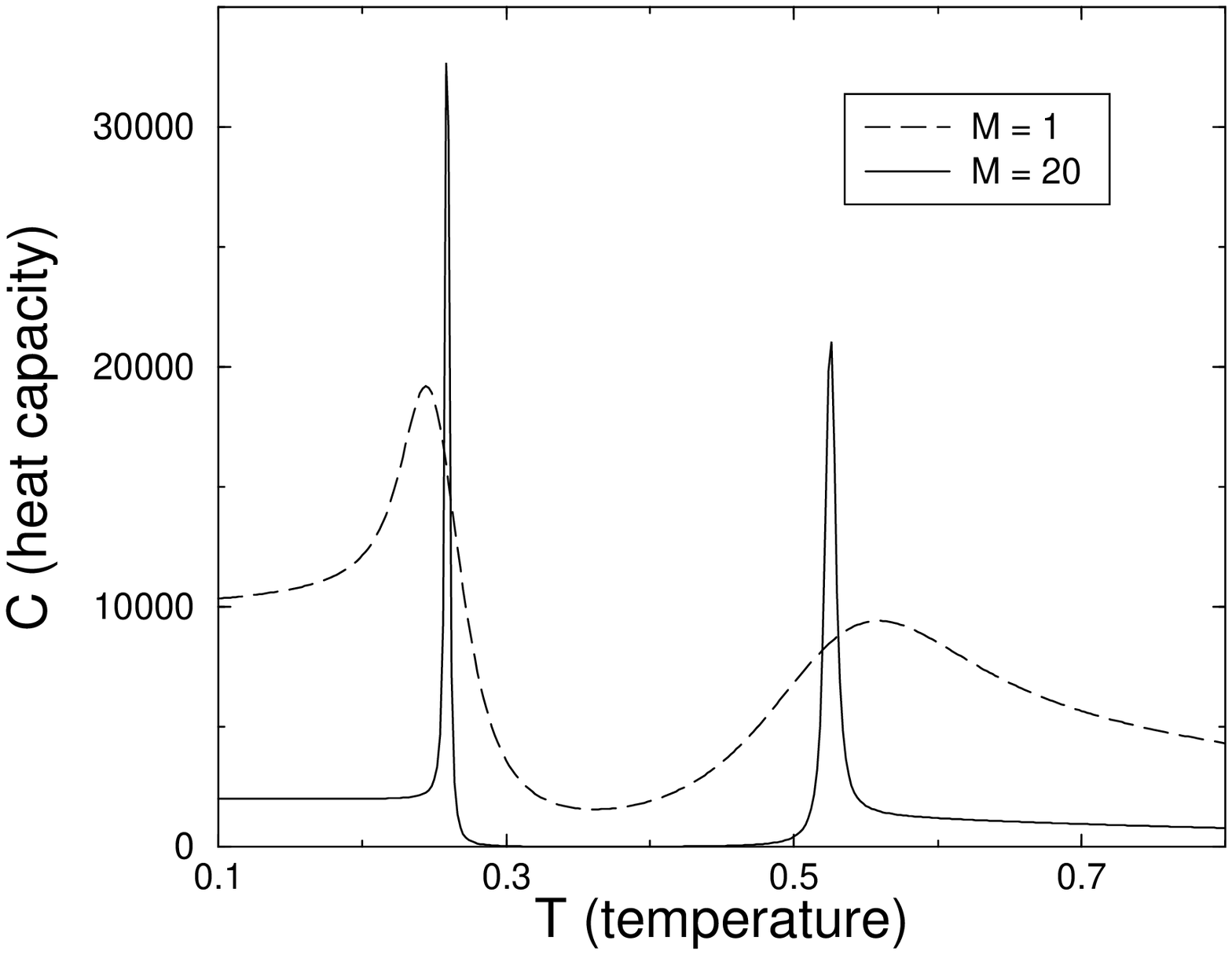,width=\linewidth}}
   \label{fig:3}
\end{figure}

The heat capacity $C(T)$ in Fig.\,\ref{fig:3} shows two characteristic peaks. Calculating the order parameter $n$, reveals that the protein is essentially unfolded in the hot and cold temperature regions. This is notable, because as earlier mentioned hot and cold unfolding is a common feature of small globular proteins. It makes sense that the protein is unfolded at low temperatures because this is a question of energy minimizing. Increasing temperature implies folding, regarded as a compromise between entropy and energy. Further increase in temperature shakes the protein, whereupon it eventually unfolds, {\it i.e.} the residual entropy of the polypeptide chain dominates in the  Gibbs free energy. It is interesting to note that the temperatures for the intersection $r=1$ for in Fig.\,\ref{fig:2} corresponds to the transition temperatures for the heat capacity in Fig.\,\ref{fig:3} for $M=20$. The heat capacity for $M=1$ is somewhat smeared out, implying a slightly broader separation between the cold and hot unfolding peaks. 

\begin{figure}
   \caption{}
   \vspace{2cm}
   \centering{\epsfig{figure=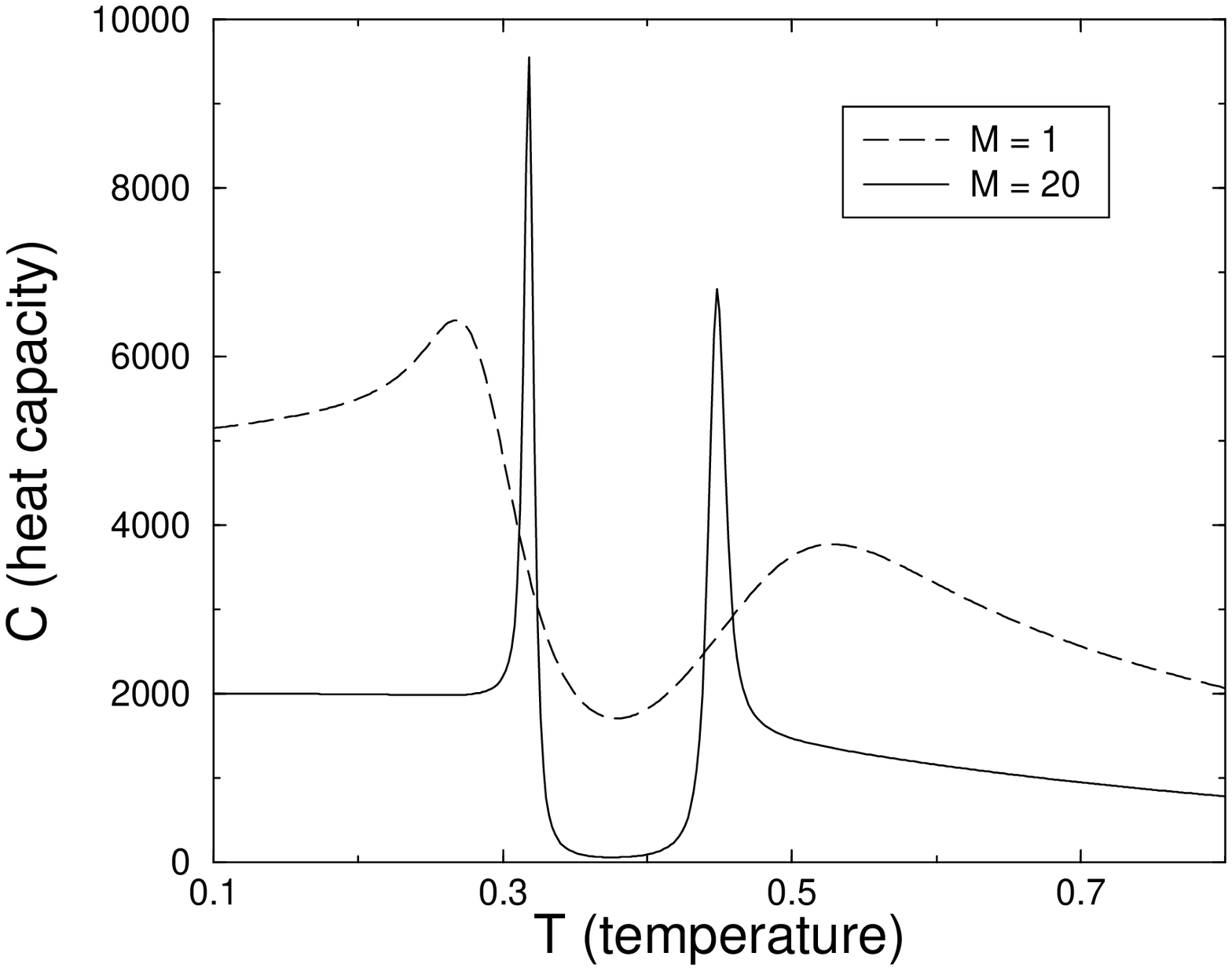,width=\linewidth}}
   \label{fig:4}
\end{figure}

\begin{figure}
   \caption{}
   \vspace{2cm}
   \centering{\epsfig{figure=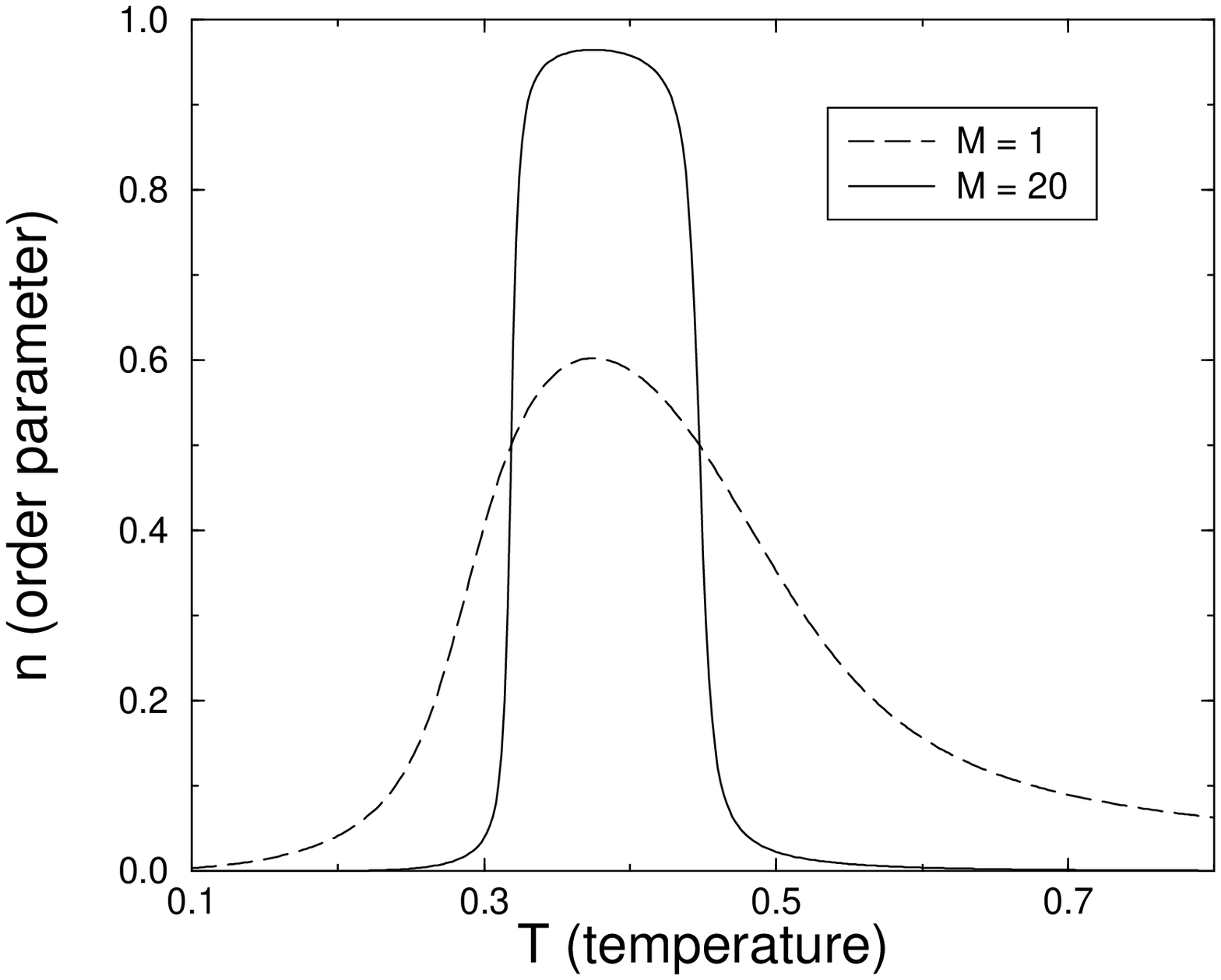,width=\linewidth}}
   \label{fig:5}
\end{figure}

Although the temperature in our model is rescaled it may be important that the relative difference between the tops in the heat capacity:\\
 $(T(\text{top } 2)-T(\text{top } 1))\,/\,T(\text{top } 2)$ corresponds to experimental data, where a typical value is $0.1-0.3$ depending on the chemical potential \cite{Privalov90}. In order to make the separation between the peaks smaller in our model, we can either decrease $\mu$ or $a$, or decrease both $\mu$ and $a$. In Fig.\,\ref{fig:4} the value of $\mu =0.635$ is slightly decreased compared to Fig.\,\ref{fig:3} where $\mu =0.65$. Obviously this results in a smaller peak separation. The order parameter $n$ in Fig.\,\ref{fig:5} shows that for $M=1$ the protein is only partly folded between the two transition temperatures, while for $M=20$ the protein is nearly completely folded. This fact suggests that for a fixed system size $N$, several water molecules per unfolded node ($M\gg 1$) is important in order to get a more realistic separation between the two peaks in the heat capacity. 

\subsection{Two level water interaction energy}
\label{sub:31}
\begin{figure}
   \caption{}
   \vspace{2cm}
   \centering{\epsfig{figure=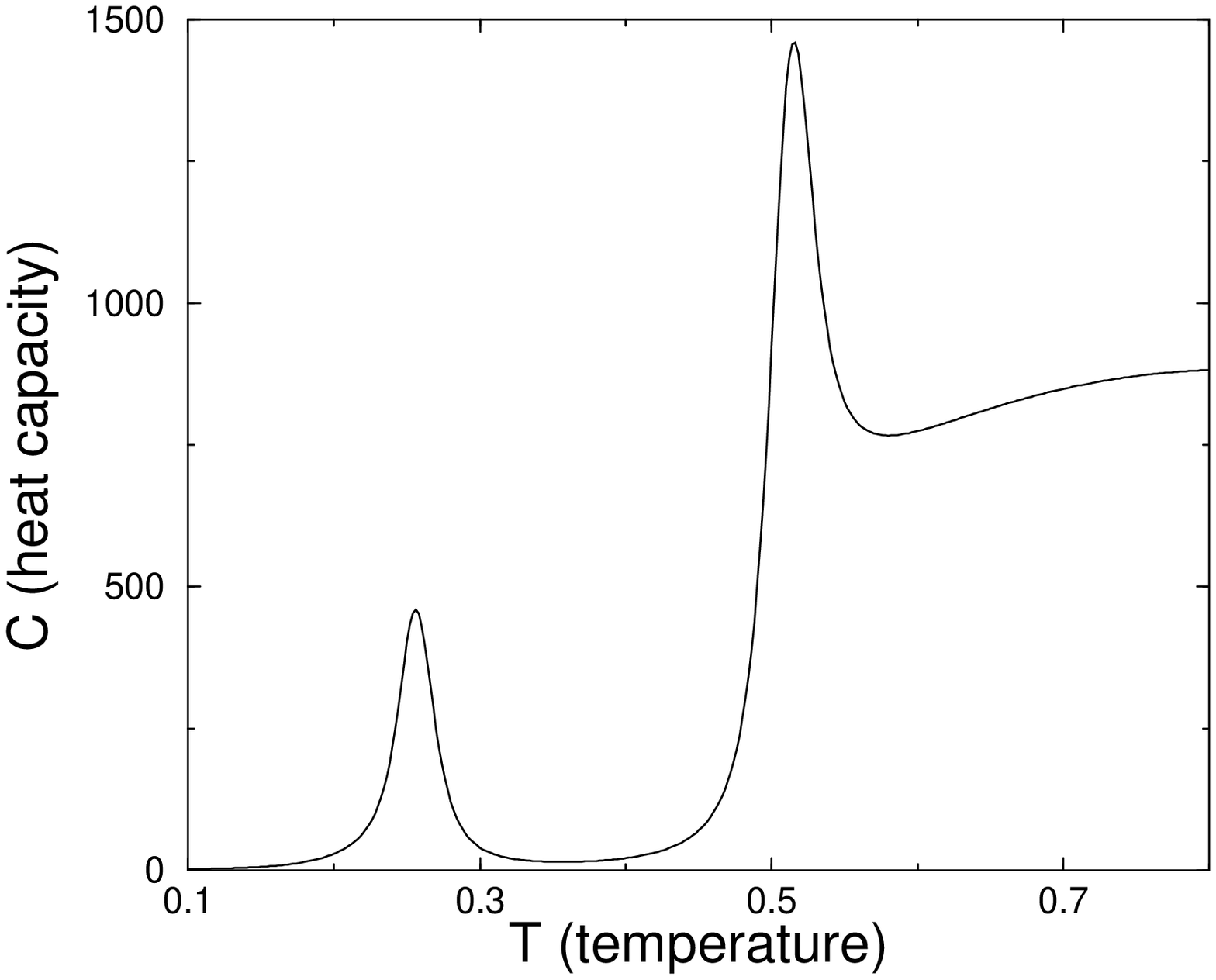,width=\linewidth}}
   \label{fig:6}
\end{figure}

Finally in this paper we will discuss the case $g=2$ for the function $r$ in Eq.\,\ref{r1}. This corresponds to an Ising spin model\,\cite{Ising25} with only two energy states per water molecule. A rearranged version of $r$ then becomes

\begin{equation}
  \label{r3}
  r=\left[ \frac{\overset{\sim}{a}\, e^{\overset{\sim}{\mu} \beta}}
                {\cosh{\beta}}\right]^M \quad ,
\end{equation}
where $\overset{\sim}{a}\equiv g/f^{1/M}$ and $\overset{\sim}{\mu}
\equiv(\epsilon_0/M -\varepsilon_w+\delta/2)\, /\, (\delta/2)$. In Fig.\,\ref{fig:6}, based on $r$ in Eq.\,\ref{r3}, one sees that the warm top is higher than the cold top, which is the opposite of the situation in Figs.\,\ref{fig:3} and\,\ref{fig:4}.  This first feature corresponds better to experimental results from Privalov {\it et al.}\,\cite{Privalov90,Privalov86}. Experiments show that, for the warm unfolding transition, the heat capacity of the unfolded state is higher than for the folded state, and it has an upward slope that decreases with increasing temperature\,\cite{Makhatadze95,Privalov97,Creighton92}, with which Fig.\,\ref{fig:6} is consistent in a qualitative way. 

Although this two-level representation of water molecules gives results with interesting features, it is not a proper representation of water. But it can give a {\it clue} to a better physical model of the system, leading to the same features of interest.

\section{Conclusion}
\label{sec:4}
We have in this paper refined the protein model proposed in Refs.\,\cite{Hansen98a,Hansen99,Bakk00} by increasing the number of water molecules that can access the hydrophobic interior of the protein. The refined model exhibits both the
hot and cold unfolding transitions. We have demonstrated how the model
only contains three effective parameters, 1) binding energy of folding relative to the orientational energy of bound water, 2)
ratio of degrees of freedom between folded and unfolded protein chain and
3) the number of water molecules that can access the hydrophobic parts of
the protein interior. By increasing the number of water molecules, we have shown that the separation
between the hot and cold unfolding transition peaks in the heat capacity
curve comes closer in comparison to the earlier protein models. This is
more consistent with the experimental data. By assuming the water-protein
interactions to be two level, which is a speculative assumption, the heat
capacity peak corresponding to the cold transition becomes smaller than
the heat capacity peak corresponding to the hot transition. This is in
agreement with experimental data, and opposite to the situation found in
the earlier protein models of Refs.\,\cite{Hansen98a,Hansen99,Bakk00}.

\vspace{2cm}
\noindent
{\large\bf Acknowledgements}\\

\noindent
A. B. thanks the Norwegian Research Council for financial support. A. H. thanks NORDITA and Niels Bohr Institute for warm hospitality and support.


\vspace{2cm}
\noindent
{\large\bf Figure captions}\\

\noindent
{\bf Fig. 1.} Schematic illustration of a partly folded protein containing $i$ folded nodes and $N-i$ unfolded nodes associated with water (shadowed).\\

\noindent
{\bf Fig. 2.} The function $r(T)$ in Eq.\,\ref{r2} for a variable effective chemical potential $\mu$, $a=0.5$ and $M=1$.\\

\noindent
{\bf Fig. 3.} Heat capacity $C(T)$ for $M=1$ (scaled by a factor 50) and $M=10$ showing two characteristic peaks for cold and hot unfolding. $a=0.5$ and $\mu =0.65$.\\

\noindent
{\bf Fig. 4.} Heat capacity $C(T)$ for $M=1$ (scaled by a factor 50) and $M=20$. $a=0.5$ and $\mu =0.635$.\\

\noindent
{\bf Fig. 5.} Order parameter $n(T)$ for $a=0.5$ and $\mu =0.635$.\\

\noindent
{\bf Fig. 6.} Heat capacity $C(T)$ for $\overset{\sim}{a}=0.48, \overset{\sim}{\mu}=0.65$ and $M=20$. This plot is based on the function $r$ in Eq.\,\ref{r3}.\\
\newpage

\vspace*{1cm}
\centering{A. Bakk, J. S. H\o ye, A. Hansen and K. Sneppen}\\
\centering{\it Physical Basis and Thermodynamical Implications of a Refined Protein Model}\\

\newpage

\vspace*{1cm}
\centering{A. Bakk, J. S. H\o ye, A. Hansen and K. Sneppen}\\
\centering{\it Physical Basis and Thermodynamical Implications of a Refined Protein Model}\\

\newpage

\vspace*{1cm}
\centering{A. Bakk, J. S. H\o ye, A. Hansen and K. Sneppen}\\
\centering{\it Physical Basis and Thermodynamical Implications of a Refined Protein Model}\\

\newpage

\vspace*{1cm}
\centering{A. Bakk, J. S. H\o ye, A. Hansen and K. Sneppen}\\
\centering{\it Physical Basis and Thermodynamical Implications of a Refined Protein Model}\\

\newpage

\vspace*{1cm}
\centering{A. Bakk, J. S. H\o ye, A. Hansen and K. Sneppen}\\
\centering{\it Physical Basis and Thermodynamical Implications of a Refined Protein Model}\\

\newpage

\vspace*{1cm}
\centering{A. Bakk, J. S. H\o ye, A. Hansen and K. Sneppen}\\
\centering{\it Physical Basis and Thermodynamical Implications of a Refined Protein Model}\\

\end{document}